\documentclass[preprint,eqsecnum,aps,showpacs]{revtex4}
\usepackage{graphicx}
\usepackage{latexsym}
\usepackage{amsmath}
\usepackage{amssymb}
\newcommand{\beq}{\begin{equation}}
\newcommand{\eeq}{\end{equation}}
\newcommand{\bea}{\begin{eqnarray}}
\newcommand{\eea}{\end{eqnarray}}
\newcommand{\bfig}{\begin{figure}}
\newcommand{\efig}{\end{figure}}
\newcommand{\nno}{\nonumber}

\newcommand{\sech}{\hbox{sech}}

\newcommand{\dis}{\displaystyle}

\begin{document}
\title{\Large \bf Chiral fermions in a spacetime with multiple warping}
\author{Ratna Koley${}^{}$\footnote{E-mail: tprk@iacs.res.in}, 
Joydip Mitra${}^{}$ \footnote{E-mail: tpjm@iacs.res.in} and  
Soumitra SenGupta${}^{}$ \footnote{E-mail: tpssg@iacs.res.in}}
\affiliation{ Department of Theoretical Physics and Centre for
Theoretical Sciences,\\
Indian Association for the Cultivation of Science,\\
Kolkata - 700 032, India}

\begin{abstract}
In a six dimensional brane world model with multiple $S_1/Z_2$ warping, two of the 4-branes at the boundaries
have coordinate dependent brane tensions in order to implement the orbifolded boundary conditions consistently.
Such brane tension is shown to be equivalent to  a scalar field distribution on the brane \cite{dcssg}.
We show that in such a scenario a masslesss left chiral fermion on the 4-brane localizes naturally on the standard model 3-brane
located at one edge of the compact manifold while the massless right chiral fermion wave function as well as
the wave functions for the massive fermion modes peak away from this brane. 
This offers a mechanism of obtaining massless chiral fermion with only one of its chiral component present in our brane. 
\end{abstract}

\maketitle

%%%%%%%%%%%%%%%%%%%%%%%%%%%%%%%%%%%%%%%%%%%%%%%%%%%%%%%%%%%%%%%%%%%%%%%%%%%%%%%%%%%%%%%%%%%%%%%%%%%%%%%%%
\section{Introduction}

Warped extra dimensional models have been studied extensively 
during the last few years ever since Randall and Sundrum proposed the two-brane model to 
resolve the gauge hierarchy problem in the standard model of 
elementary particles \cite{RS}. It has been  shown by many that the warped geometries have additional 
consequences in particle phenomenology and cosmology over and 
above the hierarchy issue \cite{phenom,horava,lykken,cohen, shiu,nelson,chodos,dienes}. 
Also in recent times several extensions of the RS model have been proposed with more than one 
extra dimension~\cite{rshigh}. Most of these models consider several 
independent $S_{1}/Z_{2}$ orbifolds along with a four dimensional Minkowski space-time~\cite{rshigh}.

Recently, Choudhury and SenGupta have proposed an alternative scenario~\cite{dcssg} where 
the warped compact dimensions get further warped by a series of successive warping leading to 
multiply warped spacetime with various p-branes sitting at the different 
orbifold fixed points satisfying appropriate boundary conditions. In this scenario 
the lower dimensional branes including the standard model 3-brane exist at the intersection edges of the higher 
dimensional branes. The resulting geometry of the multiply warped D dimensional spacetime is given 
by : $M^{1, D-1} \rightarrow \left\{ [M^{1,3} \times S^1/Z_2] \times S^1/Z_2 \right\} \times \cdots$, with $(D - 4)$ such 
warped directions. It has been argued that this multiply warped spacetime gives rise to interesting phenomenology and offers
 a possible explanation of the small mass splitting among the standard 
model fermions~\cite{dcssg}. One of the interesting characteristics of such a model 
is the bulk coordinate dependence of the higher dimensional brane tensions. Such a coordinate dependent 
brane tension is shown to be equivalent to a scalar field distribution on the higher dimensional brane which constitute the bulk for the
3-branes located at the intersection edges of these higher dimensional brane \cite{dcssg}. 
While such a scalar field distribution may have several interesting phenomenological significance for the TeV brane 
Physics, in this work we aim to study the localization issue of the standard model fermions on the visible brane, in 
particular the occurrence of massless chiral modes of fermions. In the context of the 
original RS two-brane model it has been shown that fermion localization does take place 
on the TeV brane and the standard model fermions are confined on the negative tension brane\cite{chang}. For 
Randall-Sundrum single brane model 
however one assumes the existence of a bulk scalar field which couples with the bulk fermions~\cite{ratna}. For an appropriate 
choice for this coupling, one finds that the fermion wave function can be localized on the TeV/standard model brane.

For a six dimensional multiply warped spacetime  a scalar field distribution however exists naturally in the form of a 
coordinate dependent brane tension at the two walls ( 4-branes ). This emerges naturally from the requirement of the orbifolded 
boundary conditions along the two compact directions \cite{dcssg}. In this work we first assume that a 6-dimensional  
bulk fermion is localized on the 4-brane ( i.e the 5D wall) because of the mechanism described in~\cite{chang}. For 
such five dimensional fermions however no chirality can be defined.
These five dimensional fermions can now be further localized on 
our TeV brane through the naturally occurring coordinate dependent brane tension 
which is equivalent to a scalar field distribution on the 4-brane. 
Furthermore for appropriate choice of the coupling parameter between the 
5-dimensional fermions and the scalar field distribution only the left chiral mode of 
the fermion can be localized on our TeV brane while the right handed mode gets more and more 
localized towards the other 3-brane lying at the other edge of the wall\cite{grossman}.
Furthermore the massive Kaluza-Klein (KK) modes of the fermion wave functions also tend to peak away from
the standard model 3-brane.  
This phenomena thus offers a natural explanation of the origin of chiral massless fermion mode with only one chiral component in 
our 3-brane.

The paper is organized as follows. In section II, we discuss about the multiple warped model 
briefly. Then in section III we focus on the bulk fermions residing on the 
4-branes. We examine the localization properties on the standard model 3-brane for the massless modes as well as the massive KK modes originated from the compactification in section IV and V respectively.

%%%%%%%%%%%%%%%%%%%%%%%%%%%%%%%%%%%%%%%%%%%%%%%%%%%%%%%%%%%%%%%%%%%%%%%%%%%%%%%%%%%%%%%%%%%%%%%%%%

\section{the model}
We consider the simplest model of a (5+1) dimensional anti de-Sitter bulk where both the extra dimensions are compactified in succession 
on circles with $Z_2$ orbifoldings. In this set up the solution of the Einstein equation gives rise to a doubly 
warped spacetime~\cite{dcssg} given by the following metric, 
\beq
ds^2 = b^2(z)[a^2(y)\eta_{\mu\nu}dx^{\mu}dx^{\nu} + R^2_y dy^2] + r^2_z dz^2 
\eeq
where the non-compact directions are expressed by $x^\mu \,
(\mu = 0,3)$ and the orbifolded compact directions are denoted by the angular
coordinates $y$ and $z$ respectively with $R_y$ and $r_z$ as respective moduli. Since orbifolding, in general, requires a 
localized concentration of energy, four
4-branes ($4+1$ dimensional objects) are introduced at the orbifold fixed ``points'',
namely $y = 0, \pi$ and $z = 0, \pi$.
The total bulk-brane action is given by 
%
%\beq
\bea
S =  \dis \int {d^4 x} \, {d y} \, {d z} \, 
          \left [ \sqrt{-g_6} \;  \left(R_6 - \Lambda \right) +  \sqrt{-g_5} \left [ V_1 \, \delta(y) + V_2 \, \delta( y - \pi) \right ] \right ] \\ \nno
+ \dis \int {d^4 x} \, {d y} \, {d z} \,  \left [ \sqrt{-\widetilde {g}_5} \left [
           V_3 \, \delta(z) + V_4 \, \delta(z - \pi) \right ] +  \sqrt{-g_{vis}}[{\cal L} - \hat V] \right ] \label{Action}
%S_5 & = & \dis \int {d^4 x} \, {d y} \, {d z} \, \\[1.5ex]
%    & + & \dis \int {d^4 x} \, {d y} \, {d z} \, \\[2ex]
%S_4 & = & \dis \int d^4 xdydz 
\eea
%\eeq
where the brane potential terms are, in general, $V_{1,
2} = V_{1, 2}(z) $ whereas $V_{3, 4} = V_{3, 4}(y) $. The corresponding 3-branes are
located at $(y, z) = (0,0), (0, \pi), (\pi, 0), (\pi, \pi)$. From the above 
action along with the boundary conditions the exact solution for the bulk metric can be written explicitly in the following form,
\begin{eqnarray}
ds^2 & = & \dis \frac{\cosh^2(k \, z)}{\cosh^2 (k \, \pi)} \,
  \left[ \exp\left(- 2 \, c \, |y| \right)
                \, \eta_{\mu \nu} \, d x^\mu \, d x^\nu 
     + R_y^2 \, d y^2 \right] + r_z^2 \, d z^2 \ 
\label{metric}
\end{eqnarray}
where $k = r_{z} \sqrt{\frac{- \Lambda}{10 M^4}} $ and $c = \frac{R_y k}{r_z \cosh (k \pi)} $.

The above solution with doubly orbifolded boundary conditions \cite{dcssg} results in a box-like picture of the
bulk, where the walls of the box are ($4+1$)-dimensional branes. 
%The $Z_2$ orbifoldings along the two compact coordinates puts stringent conditions on the brane tensions. 
Four ($3+1$) dimensional branes are formed at the four edges of the intersecting 4-branes. 
Our standard model 3-brane is identified with one of the four
edges ( at $y=\pi$,~$z=0$ ) by requiring the desired TeV scale while the Planck scale brane
resides at another edge. The other two edges correspond to two more
$3+1$ dimensional branes with the intermediate energy scales lying very close to TeV for one brane and close to 
Planck scale for the other. This feature results from a hierarchially different warping in the two 
compact directions. While warping in one direction is large, the other is necessarily small \cite{dcssg}. 

As discussed earlier that an important feature, relevant for this work in the multiply warped scenario is that 
the $Z_2$ orbifoldings gives rise to coordinate-dependent
brane tensions on two 4-branes which are equivalent to a scalar field distribution on the respective branes. 
Note that in this scenario the coordinate dependent brane tension effectively plays the role of a bulk scalar field 
which we need not to put in by hand but appears naturally from the requirement of orbifolded boundary conditions along
the two internal compact directions.
The (4+1) dimensional branes 
sitting at $y=0$ and $y=\pi$ have the following brane tensions,
\beq
V_1(z) \vert_{y = 0} = 8 M^2 \sqrt{\frac{-\Lambda}{10}} ~{\sech}(k\, z) = V_{0}~\sech (k\, z) = - V_2(z) \vert_{y = \pi}
\eeq

As the standard model Tev 3-brane is located at $y = \pi$, $z = 0$
we are therefore particularly interested in studying the localization of the 
bulk fermions which are residing on the 4-branes located at $y = \pi$. 
Note that the 4-brane now defines the bulk for the 3-branes located at (y, z) = ($\pi$, 0) and ($\pi$, $\pi$). 
It is expected that the coordinate dependent brane tension  $V_{2}(z)$ will play an important role in the behavior of the bulk fermions
for an appropriate coupling between them.

%%%%%%%%%%%%%%%%%%%%%%%%%%%%%%%%%%%%%%%%%%%%%%%%%%%%%%%%%%%%%%%%%%%%%%%%%%%%%%%%%%%%%%%%%%%%%%%%%%%%%%%%%
\section{Fermion localization}
While addressing the issue of fermion localization in a single brane RS model, it was found that one needs to introduce
a bulk scalar field to localize fermions on the brane where gravity is localized~\cite{oda,kogan}.
On the contrary in a RS 2-brane model it has been explicitly shown in~\cite{chang} that
the exponential warp factor leading to scale hierarchy between the two branes 
causes the 5D fermions to get localized naturally on the negative tension brane i.e on the TeV brane. Here in the six dimensional model, the warp factor along the $y$ direction is given by the usual RS warp factor $e^{-cy}$. Thus a large warping takes place between
the two 4-branes located at $y = 0$ and $ y = \pi$ leading to the localization of a six dimensional bulk fermion on the 4-brane located at $y = \pi$ which has a negative tension ( just as found in ~\cite{chang}). However in this case the brane tension being
$z$-coordinate dependent (equivalent to a scalar field distribution),  
we examine the role of such brane tension in localizing the fermion further on the TeV 
3-brane ( located at $z = 0$ and $y = \pi$) for the two different chiral states.    

The metric of the 4+1 dimensional brane at $y = \pi$ is given by 
\beq 
ds^2 = B^2(z) [ \eta_{\mu\nu} dx^{\mu} dx^{\nu}]+ dz^2 
    = B_{\pi}^2 \cosh^2( k z)[ \eta_{\mu\nu} dx^{\mu} dx^{\nu}]+ r^2_{z} dz^2 
\eeq
where, $B_{\pi} = \frac {exp(-c \pi)} {\cosh(k \pi)}$. 
The Lagrangian for the Dirac fermions in five-dimensional space-time is given by
\beq
\sqrt{-g_5} {\cal {L}}_{Dirac} = \sqrt{-g_{5}} (\bar {\psi} i \Gamma^a D_a \psi + \eta \bar {\psi} V_2(z) \psi)
\eeq
where $g_5 = det(g_{ab})$ is the determinant of the five dimensional metric and $\eta$ measures the strength of the
coupling between the fermion and the brane tension. The 4-brane tension $V_2(z)$ has been rescaled by $M$ to achieve correct dimensionality. The curved space 
gamma matrices are represented by $\Gamma^a= \left ( \frac {1} {B(z)} \gamma^{\mu},-i\gamma^5 \right)$ 
where $\gamma^{\mu},\gamma^5$ represent four dimensional gamma metrices in chiral representation. 
The Clifford algebra $\{{\Gamma^a,\Gamma^b}\} = 2 g^{ab}$ is obeyed by curved gamma metrices. 
The covariant derivative can be calculated, using the metric and is given by,
\bea
 D_{\mu} & = & \partial_{\mu} + \frac{1}{2} \Gamma_{\mu} \Gamma^4 B'(z) \\
 D_4 & = & \partial_4
\eea
For the above mentioned set up the Dirac Lagrangian turns out to be 
\beq
 \sqrt{-g_{5}} {\cal L}_{Dirac} = B^4(z)~\bar{\psi} \left [ \frac{1}{B(z)} i \gamma^{\mu} \partial_{\mu} + \gamma^5 \left( \partial_z + 2 \frac{B'(z)}{B(z)} \right) + \eta V_2(z) \right] \psi
\eeq

Now, the five-dimensional spinor can be decomposed as $\psi(x^{\mu},z)=\psi(x^{\mu}) \xi(z)$, where $\psi(x^{\mu})$
is the projection of the 5-dimensional spinor on the 3-brane.
In the massless case, we can have definite chiral states viz. $\psi_L(x^{\mu})$ and $\psi_R(x^{\mu})$ which 
correspond to left and right chiral states in four dimension. 
The $\psi_L$ and $\psi_R$ are given by, $\psi_{L,R} = \frac{1}{2} (1 \mp \gamma^5) \psi$. 
Here $\xi$ denote the extra dimensional component of the fermion wave function. 
We then can decompose five-dimensional spinor in the following way~\cite{liu},
\beq
 \psi(x^{\mu},z)=\psi_L (x^{\mu}) \xi_L(z) + \psi_R (x^{\mu}) \xi_R(z)
\eeq
Substituting the above decomposition in the Dirac Lagrangian we obtain the equations for the fermions as,
\begin{eqnarray}
\label{fermeql}
       B(z)\left[ \partial_{z} + 2 \frac{B'(z)}{B(z)} - \eta V_2(z) \right ] \xi_L(z) & = &  m \xi_R(z) \\
       B(z) \left[ \partial_{z} + 2 \frac{B'(z)}{B(z)} + \eta V_2(z) \right ] \xi_R(z) & = & - m \xi_L(z) 
 \label{fermeqr}
\end{eqnarray}
Here we have considered that the four dimensional fermions obey the standard 
equation of motion, $i \gamma^{\mu} \partial_{\mu} \psi_{L,R} = m  \psi_{R,L}$ which in turn implies that the above 
equations will be obtained  provided the following normalization conditions are satisfied:
\begin{eqnarray}
\label{norm1}
     \int^{\pi}_{0}  B^3(z) ~ \xi^m_{L,R}(z) ~ \xi^n_{L,R}(z) dz & = & \delta_{mn}  \\
     \int^{\pi}_{0}  B^3(z) ~ \xi_L^m(z) ~ \xi_R^n(z) dz & = & 0 \label{norm2}
\end{eqnarray}
We find it to be interesting to study the localization scenario of both the massless and the massive modes of chiral fermions. After getting the exact solutions of the different modes we discuss about their phemenological implications.

\section{Massless modes}
We now consider equations (\ref{fermeql}) and (\ref{fermeqr}) for two different cases namely for zero and non-zero 
coupling between the bulk fermion and coordinate dependent brane tension. 
\subsection{Coupling constant,  $\eta = 0$}

The equations of motion for left and chiral modes become
\beq
[ \partial_z + 2 k \tanh (k z)] \xi_{L,R}(z) = 0 
\eeq
The solutions of the above equations are  
\beq
         \xi_{L,R} = N_{L,R}~ {\sech}^2 (k z)
\eeq
where $N_{L,R} = \left[ \frac{k}{2~B^3_{\pi}~ \tan^{-1}\tanh(\frac{k \pi}{2})} \right]^{\frac{1}{2}}$ is the normalization 
constant, found easily from the normalization conditions in (\ref{norm1}). 
Both $\xi_{L,R}$ are peaked at $z = 0$, exhibiting the tendency  of localization around the 3-brane 
located at ($\pi$, 0) i.e. on our Standard Model brane irrespective of their chiral states. 
\bfig[ht]
\includegraphics[height = 5cm, width = 8cm]{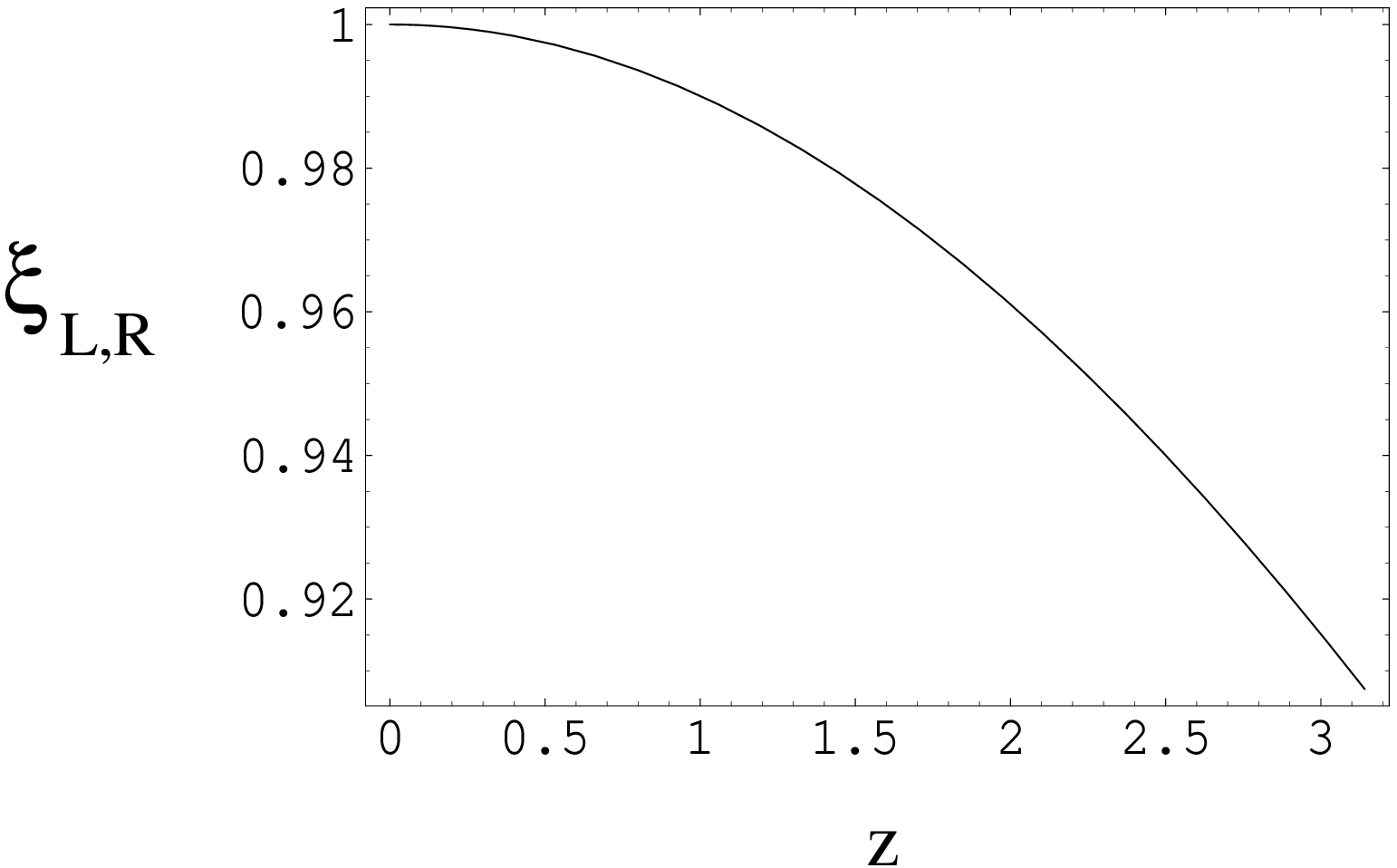}
\caption[zeromode]{Zeromodes are plotted with $k = c = 0.1$.
 Both the left and right chiral modes are peaked at z = 0}
\label{fig:etao}
\efig 
However it is apparent from the Figure (\ref{fig:etao}) that the fermion zero modes are not 
very sharply peaked at the SM brane and have a considerable extension along the extra dimension $z$. Therefore they are not 
strictly confined on the Tev brane. We shall now see a drastic change in the behavior of the wave function if we switch on the 
coupling $\eta$.  

{\subsection{Coupling constant,  $\eta \neq 0$}

Equations for the left and right chiral modes are given by 
\bea
\left [\partial_z + 2 k \tanh(k z) + \eta V_0 ~{\sech} (k z) \right] \xi_L(z) = 0 \\.
\left [\partial_z + 2 k \tanh(k z) - \eta V_0 ~{\sech} (k z) \right] \xi_R(z) = 0
\eea
The solution for the left chiral mode is
\bea
\xi_L(z)=N_L ~{\sech}^2(k z) ~\exp \left [ - \frac{2 \eta V_0}{k} \tan^{-1} \left ( \tanh \left( \frac{kz}{2} \right) \right) \right ]
\eea
where $N_L = \sqrt{\frac{2 \eta V_{0}}{B^3_{\pi} \left ( 1 - ~\exp \left [ - \frac{4 \eta V_0}{k} \tan^{-1} \left ( \tanh \left( \frac{k \pi}{2} \right) \right) \right ] \right )}}$. Figure (\ref{fig:left}) shows how the behavior of the 
left chiral mode varies with increasing values of $\eta$. We find that the these modes become 
more and more sharply peaked at the brane at $z = 0$ as the coupling $\eta$ gets stronger.
It may be noted that the maximum value of the function, $\xi^{max}_L$, increases with $\eta$ while the 
position of the maximum always remains fixed at $z = 0$. 
\bfig[ht]
\includegraphics[height = 5.5cm, width = 8.5cm]{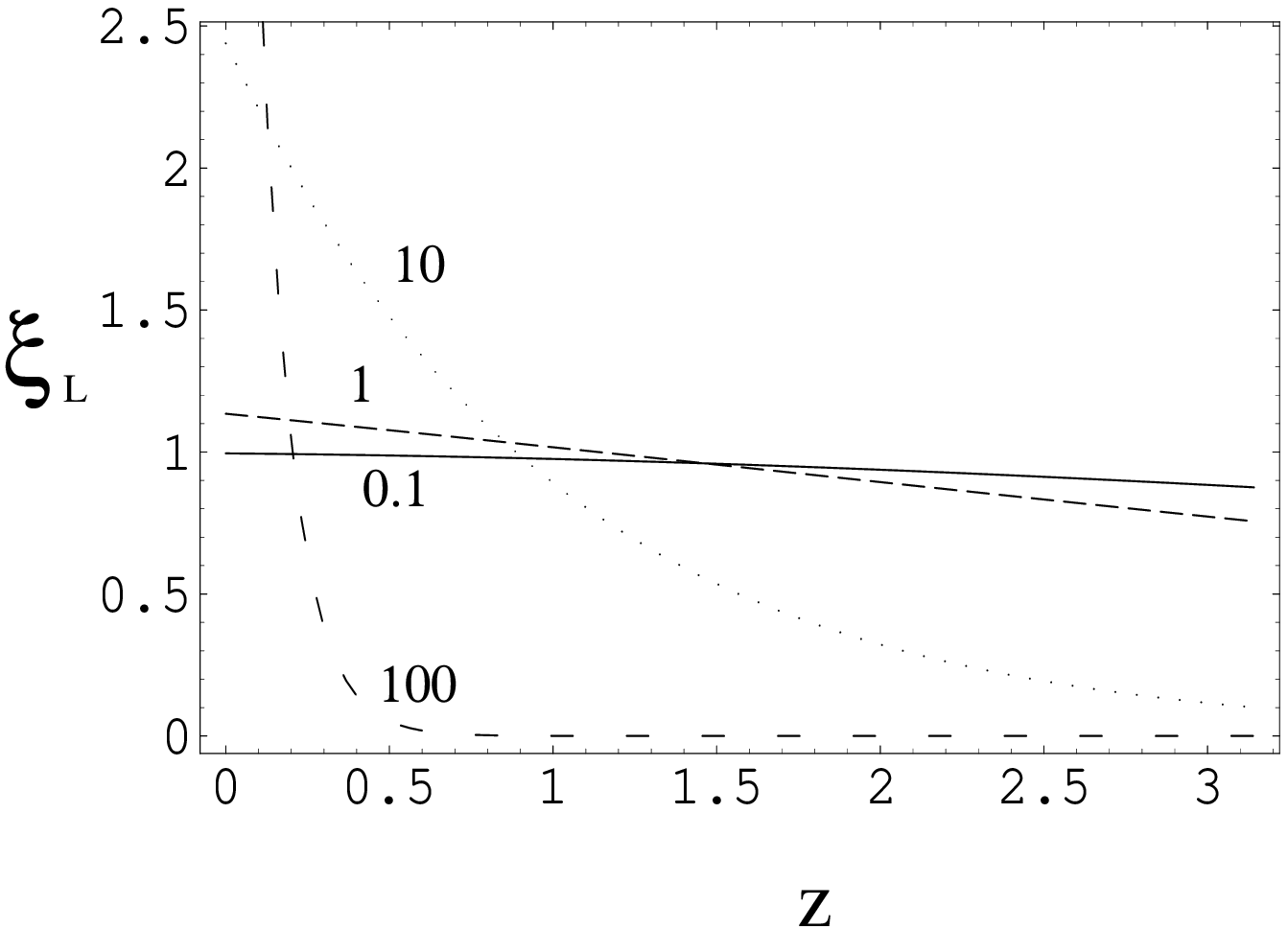}
\caption[zeromode]{We have plotted the left chiral modes for $k = c = 0.1$ for several values of $\eta$.}
\label{fig:left}
\efig
To  show explicitly the sharpness of the localization of left chiral fermions for strong coupling, 
we have studied the location of the half value of $\xi^{max}_L$ with respect to the point $z = 0$, where the peak appears.
This is plotted in Figure (\ref{fig:leftetaz}).
\bfig[ht]
\includegraphics[height = 5.5cm, width = 8.5cm]{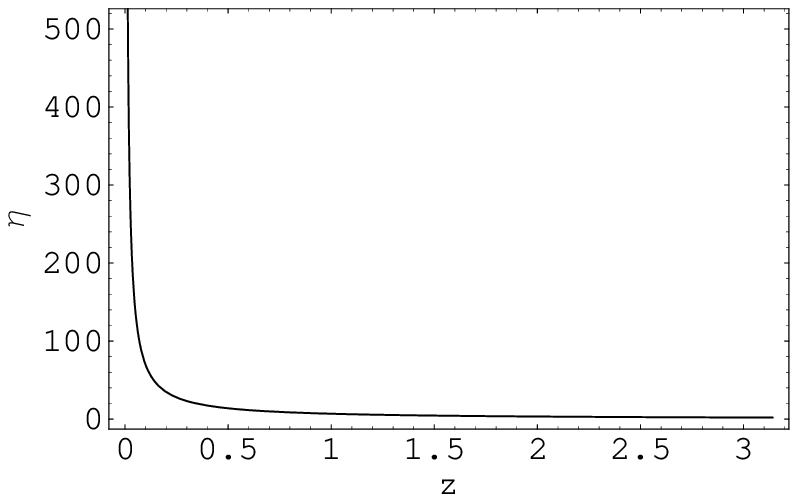}
\caption{Location of the half value of $\xi^{max}_L$ is plotted with respect to $\eta$}
\label{fig:leftetaz}
\efig
Note that as $\eta$ increases this location shifts towards the brane at $z = 0$. That implies that the left chiral mode is 
getting more more localized indicating a stronger confinement of the left chiral mode on the standard model brane.

Similarly, the solution for right mode becomes
\bea
 \xi_R(z)=N_R ~{\sech}^2(k z) ~\exp \left [ \frac{2 \eta V_0}{k} \tan^{-1} \left ( \tanh \left( \frac{kz}{2} \right) \right) \right ]
\eea
where the normalization constant $N_R = \sqrt{\frac{2 \eta V_{0}}{B^3_{\pi} \left ( - 1 + ~\exp \left [\frac{4 \eta V_0}{k} \tan^{-1} \left ( \tanh \left( \frac{k \pi}{2} \right) \right) \right ] \right )}}$.  We plot the right chiral modes in the Figure (\ref{fig:right}).
\bfig[htb]
\includegraphics[height = 5cm, width = 8cm]{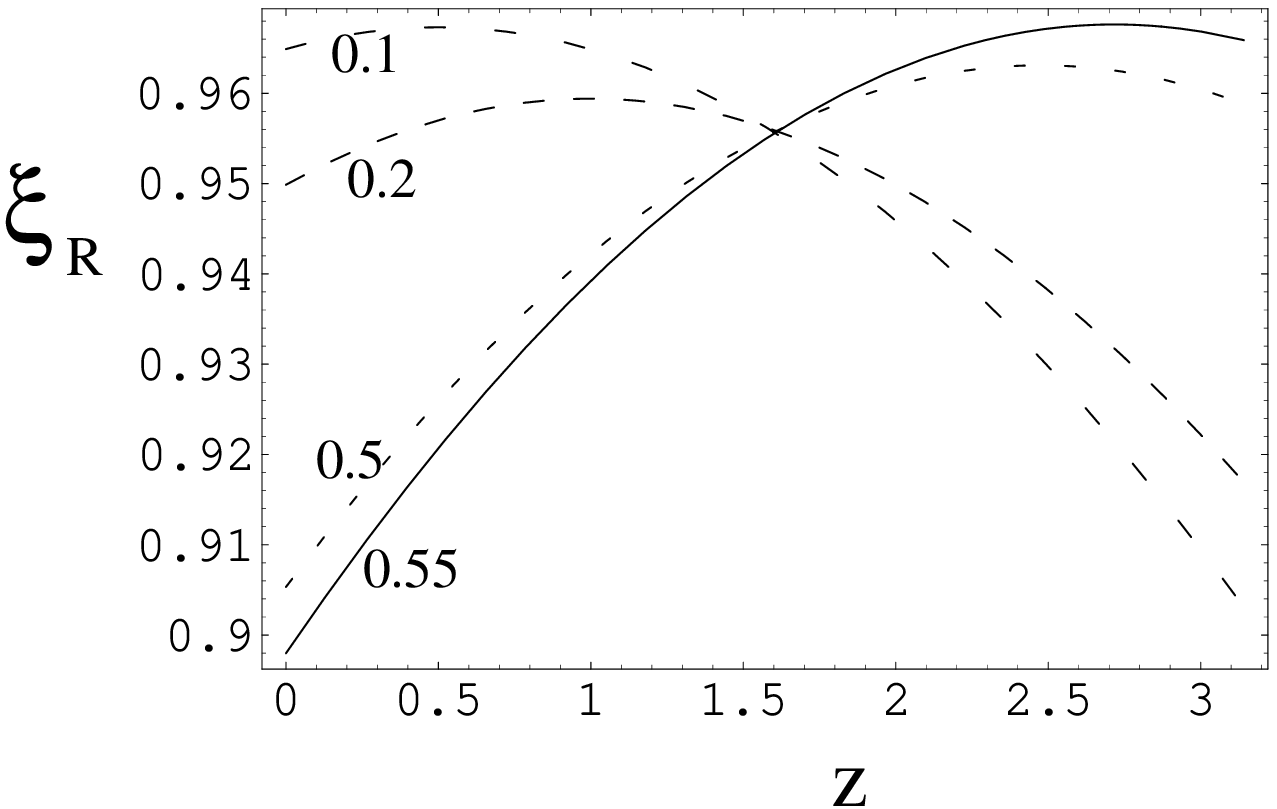}
\caption{Right chiral modes are plotted with $k = 0.2$ and $c = 0.1$ for different values of $\eta$.}
\label{fig:right}
\efig
From the figure (\ref{fig:right}) we see that the maximum value of the right mode 
depends on $\eta$. As $\eta$ increases, the z value at which $\xi_R$ becomes maximum shifts away from the brane at $z=0$. 
Further, as $\eta$ increases, the maximum value of the right mode first decreases with increasing $\eta$ and then it increases. 
We have shown this in figure (\ref{fig:rightmax}) 
\bfig[htb]
\includegraphics[height = 5cm, width = 8cm]{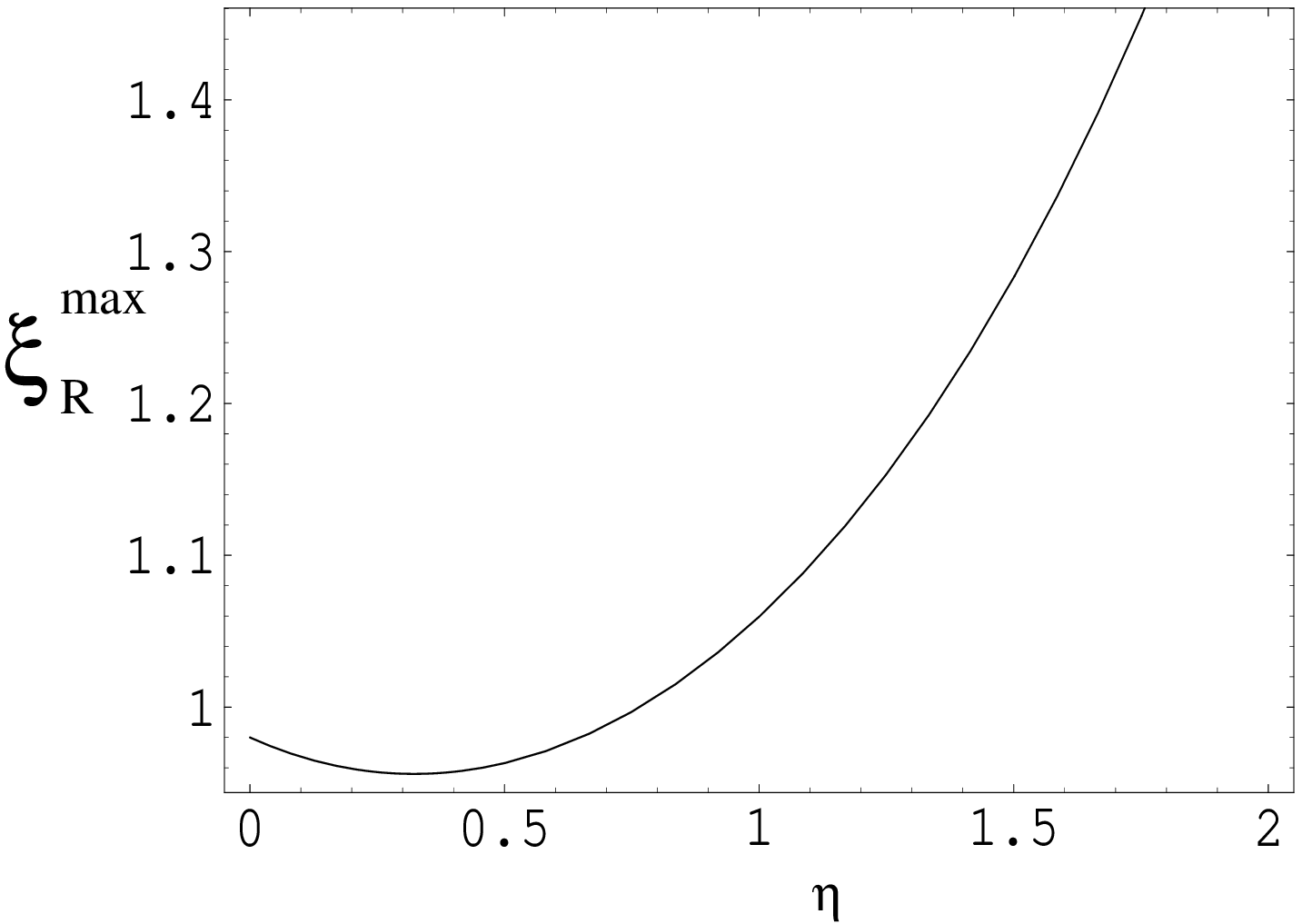}
\caption{Maximum value of $\xi_R$ plotted with respect to $\eta$.}
\label{fig:rightmax}
\efig
This clearly depicts that with increasing value of $\eta$ the left chiral mode becomes more and more localized on our standard model
brane, the right chiral mode on the other hand peaks further away from the SM brane and gets delocalized.

\section{Massive mode}
From the discussion of the massless modes , we have seen that depending on the choice of coupling parameter the 
left mode gets localized at the SM brane whereas the right mode shifts away from the brane. Of course, when 
this parameter is zero, both left and right mode solutions become identical and both modes get localized at the SM brane. In this section we turn our attention to the Kaluza-Klein tower of fermions. For an axisymmetric warped brane solution
in 6D minimal gauged supergravity it has been shown that the entire KK tower gets 
localized on the neagtive tension brane \cite{param}. However, the codimension two defects allow the KK mass gap to remain finite even in the infinite volume limit keeping the modes hidden from present day experiments. 

Let us now find out what happens to the massive KK modes in our case.  
From  the equations (\ref{fermeql}) and (\ref{fermeqr}), using 
the rescaling  $\xi_{L,R}=e^{-5 f/2} \tilde\xi_{L,R}$, we find that both $\xi_{L}$ and $\xi_{R}$ satisify the same equation which is 
given as,
\begin{equation}
\label{masmod}
\tilde\xi_{L,R}''(z)+\left[-\frac{k^2}{4}+\sech^2(kz)\left(\frac{m^2}{B_\pi^2}-\frac{k^2}{4}-\eta^2 V_0^2 \right)\right]\tilde\xi_{L,R}(z)=0
\end{equation}
As the massive states are no longer chiral we therefore subesequently drop the indices $L$ and $R$ from the
wave function and express it as $\xi(z)$. The equation of the massive modes given in (\ref{masmod}) can be reduced to 
an effective Schr\"{o}dinger equation problem where the KK modes experience an effective potential having the following form. 
\begin{equation}
U_{eff}= - \left(\frac{m^2}{B_\pi^2}-\frac{k^2}{4}-\eta^2 V_0^2 \right)\sech^2(kz)
\end{equation}

 Note that the form of the above potential is like the P\"{o}sch-Teller potential. 
The exact solutions of the massive modes can therefore be written as 
\begin{equation}
{\xi}(z)=\sech^3(k z) ~ ^2F_1\left[\epsilon-s,\epsilon+s+1,\epsilon+1,\frac{1}{2}(1-\tanh(k z))\right]
\end{equation}
where
\begin{equation}
\begin{split}
 &\epsilon=\frac{1}{2} \\
&s=\frac{1}{2}\left[-1+\frac{1}{k}\sqrt{k^2+4\left(\frac{m_n^2}{B_\pi^2}-\frac{k^2}{4}-\eta^2 V_0^2 \right)}\right]
\end{split}
\end{equation}

The mass spectrum can be obtained easily from the requirement that the wave function must be well behaved on the brane.
The possible values of the masses for these modes are found to be,
\begin{equation}
m_n^2={B_\pi^2} [k^2(n^2+2 n +1)+\eta^2 V_0^2]
\end{equation}
where n=1,2,3,....\\
It can be clearly seen from the above expression that the mass squared gap 
depends linearly on n which is given as,
\begin{equation}
\Delta m_n^2={B_\pi^2} k^2[2 n +3]
\end{equation}
Now plugging in the value of ${B_\pi^2}$ in the above expression and noting that  $k$ and $V_0$ $\sim M_P$, we find $m_n \sim$ TeV.
Thus all the massive modes have mass of the order of TeV. This apparently raises hope to find signatures of such
modes in the forthcoming TeV scale experiments at LHC. 
To address this issue we now explore whether these massive modes are localized on the SM brane. 
We draw the behaviour of some wave functions below. 
\bfig[htb]
\includegraphics[height = 5.5cm, width = 7cm]{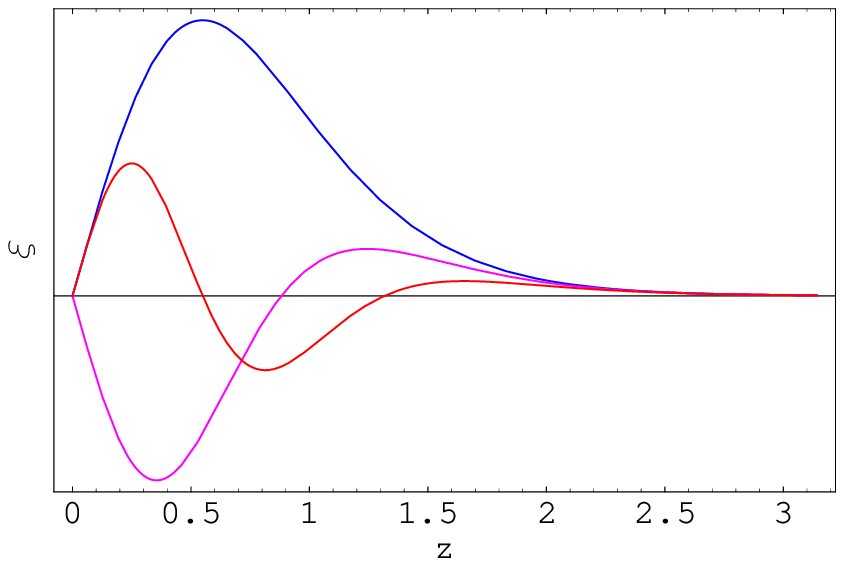}
\caption{Different modes of massive fermions has been plotted for n = 1, 3 and 5. }
\label{fig:massive}
\efig
It is clearly depicted in Fig.(\ref{fig:massive}) that the wave functions of all such massive fermion
modes peak away from the standard model (TeV) brane belying all hopes  to find their signature     
on the TeV brane.  

\section{conclusion}
Extending the earlier work \cite{dcssg}, where the 4-brane tension for the two branes at $y = 0$ and $y = \pi$ were shown to be
dependent on the orbifolded co-ordinate $z$, we have shown that such a brane tension actually plays the role of a scalar field
distribution and help to localize one of the chiral modes on the TeV 3-brane. Thus one does not need to invoke some external scalar field
by hand to achieve the localization. Thus the consistency requirement of the theory itself provides a mechanism for chirality
preferential localization. The exact dependence of the wavefunction for the two different chiral modes have been shown with respect
to their coupling with the equivalent scalar field distribution originated from the brane tension. It is found that with increasing
strength of this coupling we get the desired feature of localization of the left chiral mode on our brane while the right chiral mode
peaks away from us. In addition we have also shown that all the massive fermion KK modes have masses of the order of TeV but
the wave functions for these massive modes are not localised on the TeV 3-brane making them imperceptible 
on the TeV brane. This work therefore offers a mechanism to localize 
only the massless fermions  with a definite chirality on the visible 3-brane through multiple warping 
in a higher dimensional space-time.    

\acknowledgements{JM acknowledges Council for Scientific and Industrial Research, Govt. of India for providing financial support.}


\begin{thebibliography}{99}

\bibitem{dcssg} D. Choudhury and S. SenGupta, Phys.Rev.{ \bf D76}, 064030 (2007). 

\bibitem{RS} L. Randall and R. Sundrum, Phys. Rev. Lett. {\bf 83}, 3370 (1999);
{\it ibid} Phys. Rev. Lett. {\bf 83}, 4690 (1999).

\bibitem{phenom} H.~Davoudiasl, J.~L.~Hewett and T.~G.~Rizzo,
  Phys.\ Rev.\ Lett.\  {\bf 84}, 2080 (2000); W.~D.~Goldberger and M.~B.~Wise,
  %``Phenomenology of a stabilized modulus,''
  Phys.\ Lett.\  B {\bf 475}, 275 (2000); H. Davoudiasl and T. G. Rizzo and J. L. Hewett, Phys.Rev.{ \bf D68}, 045002 (2003); C.~Csaki, C.~Grojean, J.~Hubisz, Y.~Shirman and J.~Terning,
  Phys.\ Rev.\  D {\bf 70}, 015012 (2004); A.~L.~Fitzpatrick, J.~Kaplan, L.~Randall and L.~T.~Wang, 
{ \bf JHEP} {\bf 0709}, 013 (2007)

\bibitem{dienes} R. Dienes, E. Dudas and T. Gherghetta, Phys. Lett.{\bf B436}, 55 (1998);  Z. Kakushadze and S.H.Tye, Nucl. Phys.{\bf B548},180 (1999).

\bibitem{shiu} C.~Csaki, M.~Graesser, C.~F.~Kolda and J.~Terning,
  %``Cosmology of one extra dimension with localized gravity,''
  Phys.\ Lett.\  B {\bf 462}, 34 (1999); 
P.~Kanti, I.~I.~Kogan, K.~A.~Olive and M.~Pospelov,
  %``Cosmological 3-brane solutions,''
  Phys.\ Lett.\  B {\bf 468}, 31 (1999); 
 H.~Stoica, S.~H.~H.~Tye and I.~Wasserman,
  %``Cosmology in the Randall-Sundrum brane world scenario,''
  Phys.\ Lett.\  B {\bf 482}, 205 (2000); 
N.~Chatillon, C.~Macesanu and M.~Trodden,
  %``Brane cosmology in an arbitrary number of dimensions,''
  Phys.\ Rev.\  D {\bf 74}, 124004 (2006); 
F.~Chen, J.~M.~Cline and S.~Kanno,
  %``Modified Friedmann Equation and Inflation in Warped Codimension-two
  %Braneworld,''
  Phys.\ Rev.\  D {\bf 77}, 063531 (2008)

\bibitem{horava}  P. Horava and E. Witten, Nucl. Phys.{\bf B475},94,(1996);{ \it ibid} B460 506 (1996).

\bibitem{chodos} A. Chodos and E. Poppitz, Phys. Lett. {\bf B71},119,(1999); T. Gherghetta
and M. Shaposhnikov, Phys. Rev. Lett. {\bf 85},240,(2000).


\bibitem{cohen} A. G. Cohen and D. B. Kaplan, Phys. Lett. {\bf
B470} , 52 (1999); I. Antoniadis, S. Dimopoulos and A. Giveon,
JHEP, {\bf 05}, 055 (2001); T. Multamaki and  I. Vilja, Phys.
Lett. {\bf B545}, 389 (2002); C. P. Burgess, J. M. Cline, N. R.
Constable and H. Firouzjahi, JHEP, {\bf 01}, 014 (2002).

 \bibitem{nelson} C. Csaki and Y. Shirman, Phys. Rev.{\bf D61},024008,(2000); A.E. Nelson, Phys. Rev.{\bf D63},087503,(2001).

\bibitem{lykken}J. D. Lykken, Phys. Rev. {\bf D54}, 3693, (1996);J. Lykken and L. Randall, JHEP {\bf 06} 014 (2000);


\bibitem{rshigh} S. Randjbar-Daemi and M.E. Shaposhnikov, 
Phys.Lett.{\bf B491} 329 (2000); P. Kanti, R. Madden and K.A. Olive, Phys. Rev. {\bf D64} 044021 (2001); N.Kaloper, JHEP {\bf 0504} 061 (2004); T.Gherghetta, A.Kehagias, Phys. Rev. Lett {\bf 90} 101601 (2003). 

\bibitem{chang} S.~Chang, J.~Hisano, H.~Nakano, N.~Okada and M.~Yamaguchi, Phys. Rev {\bf D62},084025 (2000)

\bibitem{ratna} B. Bajc and G. Gabadadze, Phys. Lett. {\bf B474}, 282 (2000);  I.Oda, Phys.Lett. {\bf B496},113 (2000); C.~Ringeval, P.~Peter and J.~P.~Uzan,
  %``Localization of massive fermions on the brane,''
  Phys.\ Rev.\  D {\bf 65}, 044016 (2002); R. Koley and S. Kar, Class. Quant. Grav.  {\bf 22}, 753 (2005); {\em {ibid}} Mod. Phys. Lett {\bf A20}, 363 (2005) 

\bibitem{grossman} Y.Grossman and M.Neubert Phys.Lett {\bf B474},361-371 (2000)

\bibitem{liu} Yu-Xiau Liu {\em et. al.} hep-th 0803.0098v1(2008); S. Randjbar-Daemi and M. E. Shaposhnikov, Phys. Lett. {\bf B492}, 361,(2000)

\bibitem{oda} I. Oda, Phys.Lett. {\bf B496}, 113,(2000)

\bibitem{kogan}  I.~I.~Kogan, S.~Mouslopoulos, A.~Papazoglou and G.~G.~Ross,
  %``Multi-localization in multi-brane worlds,''
  Nucl.\ Phys.\  B {\bf 615}, 191 (2001)

\bibitem{param} S.~L.~Parameswaran, S.~Randjbar-Daemi and A.~Salvio,
  %``Gauge fields, fermions and mass gaps in 6D brane worlds,''
  Nucl.\ Phys.\  B {\bf 767}, 54 (2007); 
S.~L.~Parameswaran, S.~Randjbar-Daemi and A.~Salvio,
  %``Stability and Negative Tensions in 6D Brane Worlds,''
  JHEP {\bf 0801}, 051 (2008)


\end{thebibliography}
\end{document}